\documentstyle[12pt,{caption}]{article}
\font\tenbf=cmbx10
\font\tenrm=cmr10
\font\tenit=cmti10
\font\elevenbf=cmbx10 scaled\magstep 1
\font\elevenrm=cmr10 scaled\magstep 1
 1

\renewenvironment{thebibliography}[1]
 { \elevenrm
   \begin{list}{\arabic{enumi}.}
    {\usecounter{enumi} \setlength{\parsep}{0pt}
     \setlength{\itemsep}{3pt} \settowidth{\labelwidth}{#1.}
     \sloppy
    }}{\end{list}}
\newcommand {\ignore}[1]{}

\newcommand{\bc}{\begin{center}}
\newcommand{\ec}{\end{center}}

\def\ifmath#1{\relax\ifmmode #1\else $#1$\fi}
\def\3quarter{{\textstyle{3 \over 4}}}

\def\ra{\rightarrow}

\def\lf{\leaders\hbox to 1em{\hss.\hss}\hfill}

\def\21{$SU(2) \ot U(1)$}
\def\ne{\hbox{$\nu_e$ }}
\def\nm{\hbox{$\nu_\mu$ }}
\def\nt{\hbox{$\nu_\tau$ }}

\def\Nt{\hbox{$N_\tau$ }}

\def\O{\hbox{$\cal O$ }}

        \def\etc{\hbox{\it etc. }}
\def\eg{\hbox{\it e.g., }}        
\def\etal{\hbox{\it et al., }}

\def\gau{\hbox{gauge }}

\def\sm{\hbox{standard model }}

\def\neu{\hbox{neutrino }}
\def\sa{\hbox{such as }}

\def\neus{\hbox{neutrinos }}

\def\smp{\hbox{standard model. }}

\def\eq#1{{eq. (\ref{#1})}}

\def\lsim{\raise0.3ex\hbox{$\;<$\kern-0.75em\raise-1.1ex\hbox{$\sim\;$}}}
\def\gsim{\raise0.3ex\hbox{$\;>$\kern-0.75em\raise-1.1ex\hbox{$\sim\;$}}}

\def\bel{\begin{letter}}
\def\eel{\end{letter}}
\def\beq{\begin{equation}}
\def\eeq{\end{equation}}
\def\bef{\begin{figure}}
\def\eef{\end{figure}}
\def\bet{\begin{table}}
\def\eet{\end{table}}
\def\bea{\begin{eqnarray}}
\def\ba{\begin{array}}
\def\ea{\end{array}}
\def\bi{\begin{itemize}}
\def\ei{\end{itemize}}
\def\ben{\begin{enumerate}}
\def\een{\end{enumerate}}
\def\ra{\rightarrow}
\def\ot{\otimes}

\def\eea{\end{eqnarray}}
\def\apj#1#2#3{          {\it Astrophys. J. }{\bf #1} (19#2) #3}

\def\ib#1#2#3{           {\it ibid. }{\bf #1} (19#2) #3}

\def\nps#1#2#3{          {\it Nucl. Phys. B (Proc. Suppl.) }
                         {\bf #1} (19#2) #3}
\def\np#1#2#3{           {\it Nucl. Phys. }{\bf #1} (19#2) #3}
\def\pl#1#2#3{           {\it Phys. Lett. }{\bf #1} (19#2) #3}
\def\pr#1#2#3{           {\it Phys. Rev. }{\bf #1} (19#2) #3}
\def\prep#1#2#3{         {\it Phys. Rep. }{\bf #1} (19#2) #3}
\def\prl#1#2#3{          {\it Phys. Rev. Lett. }{\bf #1} (19#2) #3}

\def\n.c.#1#2#3{         {\it Nuovo Cim. }{\bf #1} (19#2) #3}
\def\r.n.c.#1#2#3{       {\it Riv. del Nuovo Cim. }{\bf #1} (19#2) #3}

\def\mpl#1#2#3{          {\it Mod. Phys. Lett. }{\bf #1} (19#2) #3}

\def\ppnp#1#2#3{           {\it Prog. Part. Nucl. Phys. }{\bf #1} (19#2) #3}

\relax
\begin{document}
\begin{center}{{\tenbf
               BEYOND THE STANDARD MODEL IN TAU DECAYS\\}
\vglue 1.0cm
{\tenrm JOS\'E W. F. VALLE}
\footnote{E-mail VALLE at vm.ci.uv.es or 16444::VALLE}\\
\baselineskip=13pt
{\tenit Instituto de F\'{\i}sica Corpuscular - C.S.I.C.\\
Departament de F\'{\i}sica Te\`orica, Universitat de Val\`encia\\}
\baselineskip=12pt
{\tenit 46100 Burjassot, Val\`encia, SPAIN         }\\
\vglue 0.8cm
{\tenrm ABSTRACT}}
\end{center}
\vglue 0.3cm
{\rightskip=3pc
 \leftskip=3pc
 \tenrm\baselineskip=12pt
 \noindent
The tau sector may be substantially different than
predicted in the standard model. It can have lepton
flavour violating $\tau$ decays \sa
	$\tau \ra e \pi^0$ and
	$\tau \ra e \gamma$.
Weak universality constrains the allowed rates for these
decays. The maximum branching ratio achievable when these
processes are induced by singlet neutral heavy lepton (NHL)
exchange is at the $\O(10^{-6})$ level.
I also discuss two-body decays with the emission
of a superweakly interacting massless spin zero
particle, called majoron, \eg in $\tau \ra \mu + J$.
Its branching ratio can be at the level of present
sensitivities without violating any experimental data.
In both cases the underlying physics may be probed also
at LEP through related single NHL or chargino production
processes, \eg $Z \ra \Nt \nt$ or $Z \ra \chi \tau$
(\Nt is a NHL while $\chi$ denotes the lightest chargino).
Existing observations allow these rates to be large even
when the associated \nt mass is too small to be probed
with present or expected sensitivities.
\vglue 0.6cm}
{\elevenbf\noindent Introduction}
\vglue 0.3cm
\hspace{\parindent}

Although our present description of particle
physics via the standard \21 model has been
extremely successful, it leaves several open
puzzles that motivate further extensions.

One of the puzzles of the \sm is the apparent
absence of right-handed neutrinos in nature
indicated, \eg by the experimental limits on
neutrino masses and mixing. Theoretically,
there is no fundamental symmetry that
dictates the masslessness of neutrinos and it may,
in fact, be in conflict with cosmological and
astrophysical observations involving, for example,
dark matter and solar neutrinos. The former would
indicate \neu masses at a $\sim 10$ eV scale while
the latter suggests very small neutrino masses
$\sim 10^{-3}$ eV, needed to explain the solar
neutrino data via the MSW effect. One attractive
way to generate such \neu masses is through the
exchange of singlet neutral heavy leptons (seesaw
mechanism) or as radiative effects of new scalar
bosons \cite{fae}.

Another fundamental problem in electroweak physics
today is that of mass generation, which relies
on the Higgs mechanism and, in turn, implies
the existence of fundamental scalar bosons \cite{Hunters}.
It is widely believed that some stabilizing principle
- \eg supersymmetry (SUSY) - should be effective at
the electroweak scale in order to explain the stability
of this scale against quantum corrections related to
physics at superhigh scales. Unfortunately there is no
clue as to how SUSY is realized. The most popular $ansatz$
- the minimal supersymmetric standard model (MSSM)
\cite{mssm} - postulates a discrete symmetry
called R parity ($R_p$), under which all standard
model particles are even while their partners are odd.
In this ansatz neutrinos are massless, as in the
standard model. However, this choice has no
firm theoretical basis and there are interesting
SUSY theories without R parity \cite{fae}.
Here we focus on the case of spontaneous $R_p$
breaking in the \21 theory (RPSUSY), where the
breaking of R-parity is driven by {\sl isosinglet}
slepton vacuum expectation values \cite{MASI_pot3}.
In this case the associated Goldstone boson (majoron)
is mostly singlet and as a result the $Z$ does not
decay by majoron emission, in agreement with LEP
observations \cite{LEP1}.

There is a wealth of phenomena associated with
massive neutrinos or RPSUSY \cite{granada}.
Either way of extending the \sm leads to rare tau decays,
closely related with exotic tau properties, \sa
lepton universality violating couplings and/or nonzero \nt mass.
In this talk I focus on these $\tau$ number violating decays,
and their possible relation with the single production of
SUSY particles and/or the existence of new particles, \sa
new scalar bosons or NHLS. It is important to stress that
lepton flavour violating (LFV) processes may occur with sizeable
rates even when \neus are strictly massless or
extremely light. Such rare tau decays
would be accompanied by rare $Z$ decays that
could also be accessible to experiment,
as I will illustrate with examples.

\vglue 0.6cm
{\elevenbf\noindent Preliminaries}
\vglue 0.3cm
\hspace{\parindent}

There are several bounds on \neu masses that follow
from observation. The laboratory bounds may be
summarized as \cite{PDG92}
\beq
m_{\nu_e} 	\lsim 9 \: \rm{eV}, \:\:\:\:\:
m_{\nu_\mu}	\lsim 270 \: \rm{keV}, \:\:\:\:\:
m_{\nu_\tau}	\lsim 31  \: \rm{MeV}
\eeq
In addition there are limits on neutrino mass
and mixing that follow from the nonobservation of
neutrino oscillations, which I will not repeat
here \cite{PDG92}.

Apart from laboratory limits, there is a cosmological limit that
follows from considerations related to the abundance
of relic neutrinos \cite{KT}
\beq
\sum_i m_{\nu_i} \lsim 50 \: eV
\label{rho1}
\eeq
This limit only holds if \neus are stable and
there are many ways to make \neus decay in such
a way as to avoid it \cite{fae}. The models rely
on the existence of fast \neu decays involving
majoron emission \cite{CMP,V}, \eg
\beq
\nu_\tau \ra \nu_\mu + J
\label{NUJ}
\eeq
The resulting lifetime can be made sufficiently
short so that neutrino mass values as large as
present laboratory limits are fully consistent
with astrophysics and cosmology. Examples of
seesaw type models where this is possible
have been discussed in ref. \cite{V} where
typical lifetime estimates have also been given.
Another example is provided by the spontaneously
broken R parity (RPSUSY) model \cite{ROMA}. The
region of estimated \nt lifetimes allowed in this
model is shown in Fig. 1, as a function of the \nt mass.
They should be compared to the cosmological limit on
the \nt decay lifetime required in order to efficiently
suppress the relic \nt contribution. This is shown as the
solid straight line in Fig. 1. Clearly the decay lifetimes
can be shorter than required by cosmology. Moreover, since
these decays are $invisible$, they are consistent with all
astrophysical observations. If, however, the
universe is to have become matter-dominated by a redshift
of 1000 at the latest (so that fluctuations have grown by
the same factor by today), the \nt lifetime has
to be much shorter \cite{ST}, as indicated by the dashed
line in Fig. 1. Again, lifetimes below the dashed line are
possible. However, this lifetime limit is less reliable
than the one derived from the critical density, since
there is not yet an established theory for the formation
of structure in the universe. Similarly, cosmological
big-bang nucleosynthesis constraints can also be satisfied.
\bef
\vspace{9.5cm}
\label{ntdecay}
\caption{
Estimated \nt lifetime confront observational limits}
\eef

In addition to limits, observation also
provides some positive hints for neutrino masses.
These follow from cosmological, astrophysical
and laboratory observations which we now discuss.

Recent observations from COBE indicate
the existence of hot dark matter, for
which the most attractive candidate is
a massive \neu, \sa a stable \nt,
with mass larger than a few eV \cite{cobe}.
This suggests the possibility of having
observable \ne or \nm - \nt oscillations in the
laboratory. With good luck the next generation
of experiments \sa CHORUS and NOMAD at CERN
and the P803 experiment proposed at Fermilab
will probe this possibility. In addition to
\neu oscillation signatures, some models also
suggest the possible existence of rare
muon or tau decays with appreciable rates
\cite{DARK92}. The latter could be well probed
at a tau factory.

Second, the solar \neu data collected
up to now suggests the existence of \neu
conversions involving very small \neu masses
$\lsim 10^{-2}$ eV. The region of parameters
allowed by present experiments is illustrated
in Fig. 2 \cite{GALLEX}.
\bef
\vspace{12cm}
\label{msw}
\caption{Region of \neu oscillation parameters
allowed by experiment}
\eef
Here it is
interesting to remark that the recent results
of the GALLEX experiment on low energy pp \neus do not
really "eliminate" the solar \neu puzzle,
in view of the persisting deficit of high energy
\neus seen in Kamiokande and Homestake. The
astrophysical explanation of the latter data would
require not only too large a drop in the temperature
of the solar core, but would also predict wrongly
the relative degree of suppression observed in
these two experiments \cite{Smirnov_wein}.

It is worth noting that there are some hints,
albeit controversial, from recent beta decay
studies based on solid state detectors which
indicate the presence of a 17 keV \neu
\cite{norman_wein}. This would require the
existence of a decay of the type in \eq{NUJ}.
The importance that such an observation would
have justifies the effort necessary to obtain
a conclusive confirmation or refutal of this result.

Finally, there are hints from studies involving
atmospheric neutrinos, which indicate the existence
a muon deficit \cite{atm}, etc. However I will not
discuss these in this talk.

\vglue 0.6cm
{\elevenbf\noindent Rare Decays and NHLS}
\vglue 0.3cm
\hspace{\parindent}

Neutral isosinglet heavy leptons (NHLS) arise in many
extensions of the electroweak theory \cite{SST}.
They can engender decays that are exactly forbidden
in the standard model, and whose detection would
signal new physics, closely related with the properties
of the \neus and the leptonic weak interaction. For this
reason, one may argue that such processes, if nonzero,
ought to be very small. This is the typical situation
to expect when their existence is directly related to
nonzero \neu masses. However, as discussed in
\cite{BER,CP1,CP2,CERN} this suppression need not
be always present. NHLS can mediate large LFV decays,
\sa those shown in table 1.
\begin{table}
\begin{center}
\begin{math}
\begin{array}{|c|cr|} \hline
\rm channel & \rm strength & \\
\hline
\tau \ra e \gamma ,\mu \gamma &  \sim 10^{-6} & \\
\tau \ra e \pi^0 ,\mu \pi^0 &  \sim 10^{-6} & \\
\tau \ra e \eta^0 ,\mu \eta^0 &  few \times 10^{-7} & \\
\tau \ra 3e , 3 \mu , \mu \mu e, \etc &  few \times 10^{-7} & \\
\hline
Z \ra \Nt \nt &  \sim 10^{-3} & \\
Z \ra e \tau &  few \times 10^{-7} & \\
Z \ra \mu \tau &  \sim 10^{-7} & \\
\hline
\end{array}
\end{math}
\end{center}
\caption{Maximum estimated rare $\tau$ and $Z$ decay
branching ratios consistent with lepton universality.
These processes involve isosinglet neutral heavy leptons,
and can occur even when the usual neutrinos \ne,
\nm and \nt are strictly massless.}
\end{table}
As shown in Fig. 3, taken from ref. \cite{3E},
present constraints on weak universality violation
allow the corresponding decay branching ratios to be
as large as ${\cal O} (10^{-6})$.
The solid line corresponds to the attainable branching for the
decay $l \ra l_i l_j^+l_j^-$, the dashed line
corresponds to $\tau \ra \pi^0 l_i$, the dotted line
to $\tau \ra \eta l_i$ and the dash-dotted line to
$\tau \ra l_i \gamma$. Here $l_i$ denotes $e$ or $\mu$
and all possible final-state leptons have been summed
over in each case. As one can see, the most favorable
of all the $\tau$ decay channels are $\tau \ra e \gamma$
and $\tau \ra e \pi^0$, the first being dominant for lower
NHL masses in the $100\: \rm GeV - 10 \: \rm TeV$ range.
\bef
\vspace{9cm}
\caption{Maximum estimated branching ratios for
lepton flavour violating $\tau$ decays consistent
with lepton universality. These processes may
occur even if \neus are strictly massless.}
\eef
The sensitivities of the planned tau and B factories
to these decay modes have been recently studied in
ref. \cite{TTTAU} and are summarized in table \ref{conclu}.
\begin{table}
\begin{center}
\begin{displaymath}
\begin{array}{|c|c|c|}
\hline
\mbox{channel} & \mbox{tcF} & \mbox{BF} \\
\hline \hline  & & \\
\tau\rightarrow e J &
\begin{array} {ccc}   & 10^{-5}  & \mbox{(standard-optics)}\\
                      & 10^{-6}  & \mbox{  (monochromator)}
\end{array}
&  5\times 10^{-3}   \\[0.3cm]
\hline  & & \\
\tau\rightarrow \mu  J  &
\begin{array} {ccc}   & 10^{-3}  & \mbox{      }\\
                      & 10^{-4}  & \mbox{(RICH)}
\end{array}
&   5\times 10^{-3}  \\[0.3cm]
\hline & & \\
\begin{array} {c}
\tau\rightarrow e \gamma \\
\tau\rightarrow \mu \gamma
\end{array}
 &  10^{-7} & 10^{-6}\\[0.3cm]
\hline   & & \\
\tau\rightarrow \mu \mu\mu     &  &  \\
\tau\rightarrow \mu e e     & 10^{-7}   &10^{-7}  \\
\tau\rightarrow  e  \mu\mu     &     &    \\
\tau\rightarrow  e  e e     &    &    \\[0.3cm]
\hline
\end{array}
\end{displaymath}
\end{center}
\caption{Attainable limits for the branching ratios for  different $\tau$
decays in a $\tau$-charm Factory and a B Factory for one year run.
J is  a neutrino like particle (Majoron).}
\label{conclu}
\end{table}

The physics of rare $Z$ decays beautifully
complements what can be learned from the
study of rare LFV $\tau$ lepton decays.
For example, the exchange of isosinglet
neutral heavy leptons can induce new $Z$
decays \sa
\footnote{There may also be CP violation in lepton
sector, despite the fact that the physical light \neus
are strictly massless \cite{CP1}. The corresponding
decay asymmetries can be of order unity with respect
to the corresponding LFV decays \cite{CP2}. },
\beq
\begin{array}{lr}
Z \ra e + \tau \: , & Z \ra N_{\tau} + \nu_{\tau}
\end{array}
\eeq
Taking into account the constraints on the parameters
describing the leptonic weak interaction one can
estimate the attainable values for these branching
ratios given in table 1. The most copious channel
is $Z \ra \Nt \nt$, possible if the \Nt is lighter
than the $Z$ \cite{CERN}. Subsequent \Nt decays
would then give rise to large missing $p_T$ events,
called zen-events. Prompted by our suggestion,
there are now good limits on such decays from the
searches for acoplanar jets and lepton pairs from $Z$
decays at LEP \cite{opal}. The method of experimental
analysis is very similar to that used in SUSY zen-event
searches.

If $M_{\Nt}>M_Z$ then \Nt can not be directly produced at
LEP1 but can still mediate rare LFV decays \sa $Z \ra e \tau$
or $Z \ra \mu \tau$ through virtual loops. Under realistic
luminosity and experimental resolution assumptions, it is
unlikely that one will be able to see these processes at LEP
\cite{ETAU}. In contrast, the related low energy processes seem
to be within the expectations of a $\tau$ or $B$ factory.
In any case, there have been dedicated searches for flavour
violation at the $Z$ peak at LEP, and some limits have already
been obtained \cite{opal}. This example illustrates how the
search for rare decays is {\it complementary to the physics of
\neu mass per se}. For more details the reader
is referred to ref. \cite{BER,CP1,CP2,CERN,3E}.

\vglue 0.6cm
{\elevenbf\noindent Rare Decays and Supersymmetry}
\vglue 0.3cm
\hspace{\parindent}

Supersymmetric models can also produce rare lepton
flavour violating decays with detectable rates.
Here we are concerned with the situation where
supersymmetry is realized in a \21 context in such
a way that R parity is broken spontaneously at or
slightly below the TeV scale \cite{MASI_pot3}.
In such RPSUSY models one can have several
rare decay processes, \sa those of table 3.
\begin{table}
\begin{center}
\begin{math}
\begin{array}{|c|cr|} \hline
\rm channel & \rm strength & \\
\hline
\tau \ra \mu + J &  \sim 10^{-3} & \\
\tau \ra e + J &  \sim 10^{-4} & \\
\hline
Z \ra \chi \tau &  6 \times 10^{-5} & \\
%
\hline
\end{array}
\end{math}
\end{center}
\caption{Attainable branching ratios for
$\tau$ and $Z$ decays in SUSY models with
spontaneously broken R parity. Here $J$
denotes the majoron, while $\chi$ denotes
the lightest charged SUSY fermion}
\end{table}

Since in RPSUSY models also lepton number is
broken spontaneously, there is a physical Goldstone
boson, called majoron. Its existence is consistent with
recent measurements of the $invisible$ $Z$ decay width
at LEP, since the majoron is a singlet under the \21
\gau symmetry. Its existence leads to a new class of
lepton flavour violating $\tau$ decays. These involve
the emission of such weakly interacting massless
pseudoscalar majorons, which would be "seen"
only insofar as their emission would affect
the spectra of the leptons produced in $\tau$
decays.
Single majoron emission in $\tau$ decays would lead
to bumps in the final lepton energy spectrum,
at half of the $\tau$ mass in its rest frame.
The allowed emission rates have been determined by varying
the relevant parameters over reasonable ranges and imposing
all of the observational constraints \cite{NPBTAU}.
A specially important role is played by
constraints related to the flavour and/or
total lepton number violating processes \sa
those arising from negative neutrino oscillation
and neutrinoless double $\beta$ decay searches,
as well as from the failure to observe
anomalous peaks on the energy distribution of
the electrons and muons coming from decays \sa
$\pi, K \ra e \nu$ and $\pi, K \ra \mu \nu$.
Our results are shown in Fig. 4, taken from ref.
\cite{NPBTAU}. One sees that these majoron emitting
$\tau$ decay modes may easily occur at a level $\O(10^{-3})$
in branching ratio \cite{NPBTAU}, close to the present
ARGUS limit $BR(\tau \ra \mu + J) \leq 5.8 \times 10^{-3}$
\cite{SINGLE}.
\bef
\vspace{9cm}
\caption{Attainable branching ratios for lepton flavour
violating $\tau$ decays with majoron emission.}
\eef
The points in the region delimited by the solid contours
correspond to different values of $\tan\beta$, an
unknown SUSY model parameter expected to lie between 1
and $\frac{m_t}{m_b}$, the top-bottom quark mass ratio.
Similar results hold for the case of the decay $ \tau \ra e + J$.
In this case the attainable branching ratio cannot exceed
$\rm few \times 10^{-4}$, to be contrasted with the present
ARGUS limit $BR(\tau \ra e + J) \leq 3.2 \times 10^{-3}$.
The sensitivity of the planned tau factories
to this decay mode has been recently shown to
be at the $10^{-5}$ level with standard optics
and $10^{-6}$ in schemes with monocromator
(see table \ref{conclu}, taken from ref. \cite{TTTAU}).

In addition to rare $\tau$ decays, RPSUSY leads to
corresponding rare $Z$ decays, \sa
\beq
Z \ra \chi \tau
\eeq
The single production of the lightest $chargino$
in $Z$ decays would lead to striking signatures at
LEP \cite{ROMA}. As shown in Fig. 5, the allowed
branching ratio lies close to the present LEP
sensitivities.
\bef
\vspace{9cm}
\label{}
\caption{Maximum estimated branching ratios for single
chargino production in $Z$ decays, $Z \ra \chi + \tau$,
as a function of the chargino and \nt masses. Comparing
this with Fig. 4 illustrates the complementarity
of $Z$ and $\tau$ physics in the RPSUSY model}
\eef
In addition, in this model the lightest neutralino
is unstable and is therefore not necessarily
an origin of events with missing energy.
Zen events, similar to those of the
MSSM are expected say, from the usual pair
neutralino production process, where one $\chi^0$
decays visibly and the other invisibly, with the
missing energy carried by the \nt and the majoron.
The corresponding zen-event rates can be larger than
in the minimal SUSY model and their origin is also
quite different.
Moreover, these processes are intimately tied with
each other and to the nonzero value of the \nt mass.

Finally we remark that in the RPSUSY model \neus
acquire masses as a result of lepton number violation.
These masses follow an interesting pattern,
according to which the heavier \neu is the \nt,
whose mass can be very large and determines the strength
of the rare decay processes we have discussed. The
\ne and \nm are very light, with a mass difference
that can easily lie in the range where MSW \ne to \nm
conversions may provide an explanation of
solar \neu data. In fact one can go even
further, and regard the rare processes discussed
here as a possible tool to probe the physics
underlying the solar \neu deficit. It has been
shown \cite{RPMSW} that indeed the rates for
such rare decays can be used in order to
discriminate between large and small mixing
angle MSW solutions to the solar \neu problem.
For example, only in the nonadiabatic region of
small mixing one can have a branching ratio for
$\tau \ra \mu J$ $and$ $Z \ra \tau \chi$ in
excess of $10^{-5}$, as seen in Fig. 5 of ref.
\cite{RPMSW}.

\vglue 0.6cm
{\elevenbf\noindent Discussion}
\vglue 0.3cm
\hspace{\parindent}

Upcoming facilities \sa a tau factory or a $B$ factory,
can be used to probe for the possible existence of the
$\tau$ decay processes discussed here. If no signal
is found, one will obtain restrictions on the relevant
parameters of these extensions of the \smp
Even if an improvement on the \nt mass or universality
limits is obtained in the near future, it is not likely
to narrow down the rare $\tau$ and $Z$ decay possibilities
discussed here to a level where they are rendered undetectable.

Any novel physics, \sa discussed here for the $\tau$
lepton is bound to be manifest, as I have illustrated,
also at the $Z$ pole, and thus produce signals at LEP.
The search for rare $Z$ decays gives another way to test
fundamental symmetries of the standard electroweak theory,
\sa weak universality, flavour and CP conservation.
The underlying physics is the same, although the
detectability prospects may differ from those of
the corresponding low energy processes.

\vskip 5mm
{\bf \noindent Acknowledgements}
\vskip 2.3mm
I thank the organizers for a very efficient
and friendly organization and my collaborator
J. C. Rom\~{a}o for preparing Fig. 5.
\vskip 5mm

\bibliographystyle{ansrt}

\end{document}